\documentclass[12pt]{article}
\usepackage[T2A]{fontenc}
\usepackage[cp1251]{inputenc}
\usepackage[english]{babel}
\usepackage[dvips]{graphicx}
\normalsize
\usepackage{amssymb,amsmath}
\newcommand{\Teff}{T_{\rm eff}}
\newcommand{\logg}{\rm log~ g}
\newcommand{\eps}[1]{\log\varepsilon_{\rm #1}}
\newcommand{\Eexc}{$E_{\rm exc}$}
\newcommand{\kH}{$S_{\!\!\rm H}$}    %%% Note negative spaces!
\newcommand{\kms}{km\,s$^{-1}$}

\newcommand{\ang}{\rm\mathring{A}}

\begin{document}

\title{Formation of Zr I and II lines under non-LTE conditions of stellar atmospheres}
\author{A.\,Velichko$^1$\thanks{e-mail: anna@inasan.rssi.ru}, L.\,Mashonkina$^1$, H.\,Nilsson$^2$\\[2mm]
\begin{tabular}{l}
 $^1$ {\it Institut of Astronomy of RAS, Moscow, Russia}\\[2mm]
 $^2$ {\it Lund observatory, Sweden}
\end{tabular}
}
\date{}
\maketitle

\begin{abstract}
The non-local thermodynaic equilibrium (non-LTE) line formation for the two ions of zirconium is considered through a range of spectral types when the Zr abundance varies from the solar value down to [Zr/H] = $-3$.
% Zr~I-Zr~II in the atmospheres of cool stars was considered for the first time. 
The model atom was built using 148 energy levels of Zr~I, 772 levels of Zr~II, and the ground state of Zr~III. 
%The non-LTE calculations were performed for the model atmospheres with $\Teff$ = 5500~K and 6000~K, $\logg$ = 2.0 and 4.0, [M/H] = $-3, -2, -1$, and 0. 
It was shown that the main non-LTE mechnism for the minority species Zr~I is ultraviolet overionization. Non-LTE leads to systematically depleted total absorption in the Zr~I lines and positive abundance corrections, reaching to 0.33~dex for the solar metallicity models. The excited levels of Zr~II are overpopulated relative to their thermodynamic equilibrium (TE) populations in the line formation layers due to radiative pumping from the low-excitation levels. As a result, the line source function exceeds the Planck function leading to weakening the Zr~II lines and positive non-LTE abundance corrections. Such corrections grow towards lower metallicity and lower surface gravity and reach to 0.34~dex for $\logg$ = 2.0 and [M/H] = $-2$ at $\Teff$ =  5500~K. 
%levels are underpopulated in the line formation layers. Opposite, the excited levels of Zr~II are overpopulated, while the ground state and low-excitation levels keep their TE populations. Non-LTE leads to weakening both the Zr~I and Zr~II lines investigated For the Zr~II lines, they 
%The non-LTE effects are weakly sensitive to $\Teff$ variation. The non-LTE abundance corrections for Zr~I lines can reach to 0.33~dex for the solar metallicity models. 
As a test and first application of the Zr~I-Zr~II model atom, Zr 
abundance was determined for the Sun on the basis of 1D LTE model atmosphere. Lines of Zr~I and Zr~II give consistent within the error bars non-LTE abundances, while the difference in LTE abundances amounts to 0.28~dex. 
%It was shown from analysis of the solar Zr~I and Zr~II lines that the solar zirconium non-LTE abundances determined from two ionization stages agree within the error bars
The solar abundance of zirconium obtained with the MAFAGS solar model atmosphere is $\eps{Zr,\odot}$ = 2.63$\pm$0.07.
\end{abstract}
%\maketitle

\section{Introduction}

One believes that the elements beyound the iron group are produced by neutron-capture reactions. These reactions are subdivided into rapid (r) and slow (s) processes depending on the neutron flux available. The main and secondary or weak component are distinguished among the slow processes. The r-process is often associated with type-II supernova explosions. The weak and main components of the s-process can take place, respectively, in the core of high-mass ($M > 20M_\odot$) stars at the hydrostatic helium core burning stage and during the thermally-pulsing AGB phase of intermediate-mass ($2 - 4 M_\odot$) stars. According to K${\rm\ddot{a}}$ppeler~et~al. \cite{kappeler}, in high-mass stars the nuclei of only the lightest of the heavy elements, with atomic masses $A < 90$, can be produced in the s-process. Since the atomic mass of zirconium is $A$ = 91.22, it can be produced in all three types of neutron-capture reactions. The characteristic production time is different for different types of reactions. Therefore, the relative contributions from each of the processes to the abundance of a specific element changed as the Galaxy evolved. At present, the theories of the r-process and the weak component of the s-process cannot accurately predict the yields of elements. Therefore, it is very important to restore the history of heavy elements enrichment of the interstellar medium and, thus, to impose constraints on the nucleosynthesis theories on the basis of observational data.

In our previous paper \cite{yzrce}, we determined the zirconium abundance assuming the local thermodynamic equilibrium (LTE) for a sample of 52 stars belonging to three types of Galactic population - the thin disk, the thick disk, and the halo. We found a large overabundance of zirconium relative to barium, up to log(Zr/Ba) = 1.9 at [Ba/H] = $-3.8$, in the halo stars. The Zr/Ba ratio decreases towards higher barium abundance and reaches its solar value of log(Zr$_\odot$/Ba$_\odot$) = 0.41 \cite{lodders} in the thin-disk stars. If the heavy elements would synthesized only in the r-process in the early Galaxy, until the first intermediate-mass stars evolved, the ratio log(Zr/Ba) in old stars must be constant at a level of 0.33~dex \cite{arlandini}. Our observational data appeared to infer a
distinct production mechanism for the light trans-iron (Sr--Zr) and heavy
elements beyond Ba in the early Galaxy. Similar conclusions were drawn in \cite{aoki05,francois2007}.

In this paper, we want to check whether departures from LTE can affect the determined evolutionary behavior of the zirconium abundance. We develop a technique for calculating the statistical equilibrium (SE) of Zr~I-Zr~II and consider the non-local thermodynamic equilibrium (non-LTE) line formation for Zr~I and Zr~II in the atmospheres of the Sun and cool stars. 

The paper is structured as follows. In the next section, we describe the Zr~I-Zr~II model atom, the method of non-LTE calculations, and the mechanisms of departure from LTE for Zr~I and Zr~II. In Sect.\,\ref{Sun}, we analyze the solar zirconium lines, determine the solar zirconium abundance from Zr~I and Zr~II lines for various line formation scenarios, and compare our LTE results with the literature data. The non-LTE effects for Zr~I and Zr~II lines depending on stellar parameters are presented in Sect.\,\ref{NLTE_effects}. Finally, we give our conclusions.

\section{Non-LTE calculations}

\subsection{Model atom}

{\it Energy levels}. In the atmosphere of cool stars with effective temperature $\Teff \ge$ 5000~K, zirconium is largely ionized and the fraction of Zr~I atoms does not exceed a thousandth of the total number of zirconium atoms. Therefore, the Zr~I level populations are very sensitive to a change in the net ionization minus recombination rate. The Zr~II ionization stage is of our particular interest, because only Zr~II lines are detected in the metal-deficient stars.

The system of Zr~I measured levels includes the singlet, triplet, and quintet terms of the electronic configurations 4d$^2$5s$^2$; 4d$^2$5s5p; 4d$^3$nl, where nl = 5s and 5p, in total, 148 levels with excitation energy up to \Eexc\,= 4.25~eV from the NIST (http://physics.nist.gov/ PhysRefData) and VALD \cite{kupka1999} atomic data databases. The ground state of Zr~I is 4d$^2$5s$^2$~$^3$F$_2$.

\begin{figure}

%\hbox{
\resizebox{130mm}{!} {\includegraphics{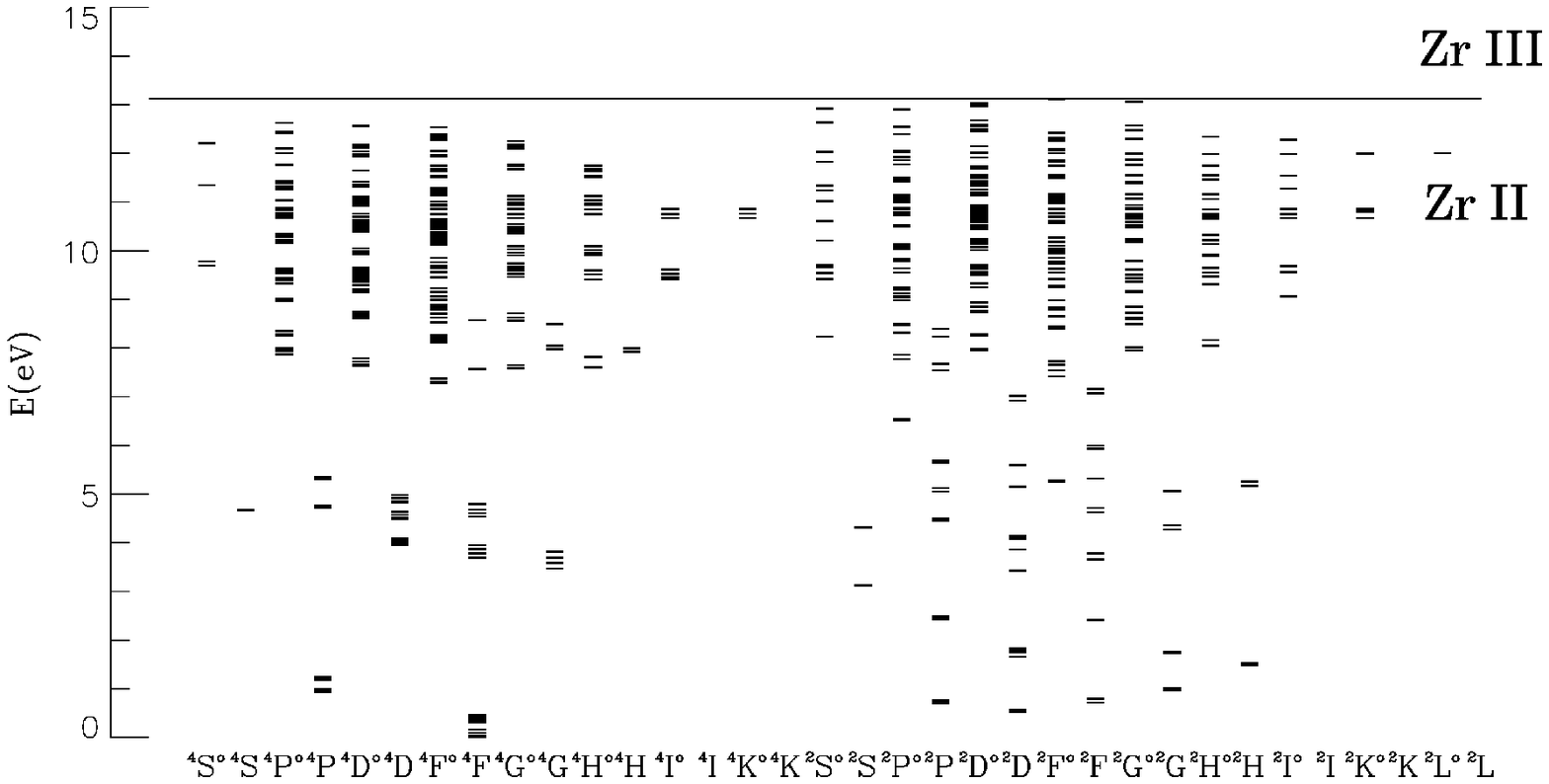}}
%}
%\hbox{
\resizebox{130mm}{!} {\includegraphics{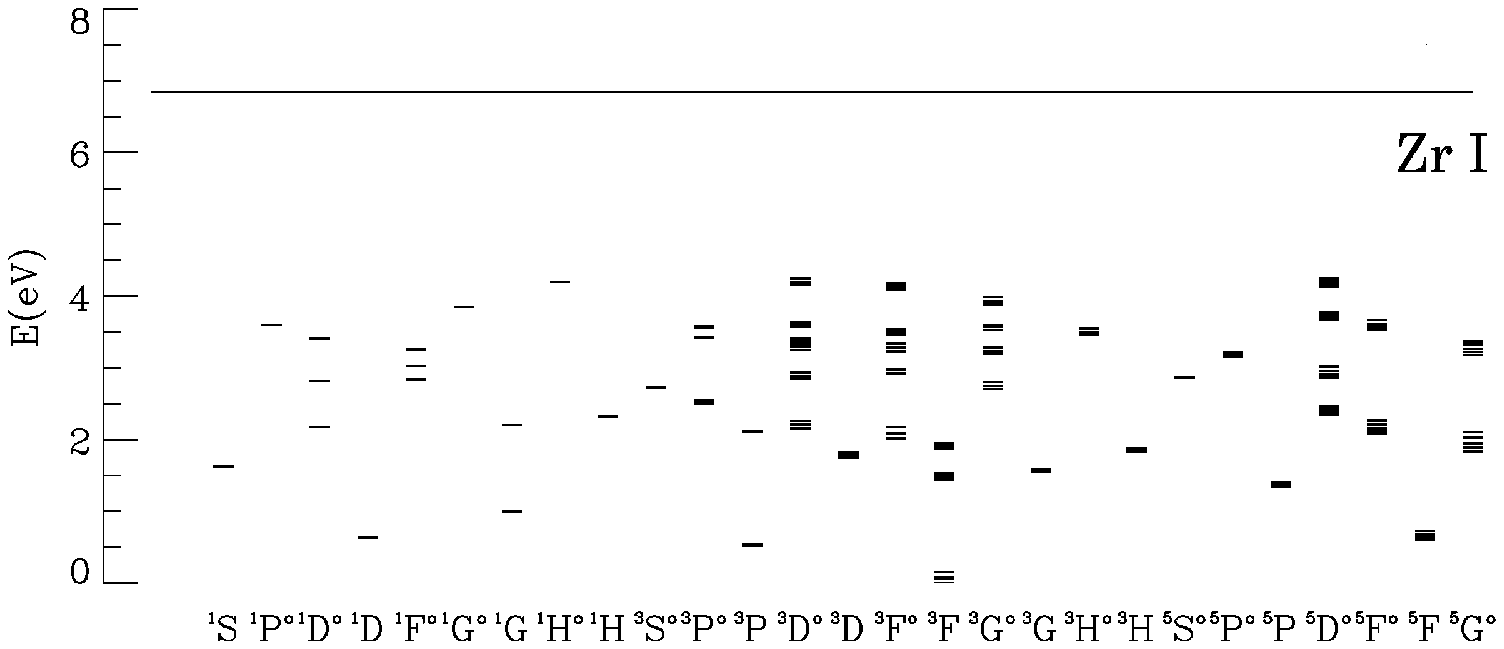}}
% }

\caption{ The term diagram of Zr~I-Zr~II as known from laboratory measurements and atomic structure calculations for Zr~II.} \label{model-at}
\end{figure}

The experimental Zr~II levels (a total of 103) provided by the NIST database and by \cite{malch} have \Eexc\,$\leq$ 8.75~eV and belong to the doublet and quartet terms of the electronic configurations 4d$^2$nl (nl = 5s, 5p), 4d5s$^2$, and 4d$^3$. The ground state of Zr~II is 4d$^2$5s~$^4\rm F_{3/2}$. We also employ 669 levels predicted in our calculations of the Zr~II atomic structure. These levels belong to the doublet and quartet terms of the following electronic configurations: 4d$^3$, 4d$^2$nl, where nl = 5p, 5d, 5f, 5g, 6s, 6d, 6p, 7s, 7d, and 7p, and 4d5snl, where nl = 4f, 5p, 5d, 6s, 6p, 6d, and 7p. The term diagram of Zr~I and Zr~II is shown in Fig.\ref{model-at}.  

In constructing the model atom, the fine structure was taken into account for all the Zr~II levels with \Eexc\,$< 5$~eV. The remaining levels with common parity and close energies were combined. 
The final model atom includes 63 levels of Zr~I, 247 levels of Zr~II, and the ground state of Zr~III.

{\it Radiative bound-bound (b-b) rates}. The system of the Zr~I levels in our model atom includes 246 permitted bound-bound transitions. The oscillator strengths ($f_{ij}$) were taken from the NIST and VALD databases. The Zr~II model atom includes 9336 radiative $b-b$ transitions. For 1070 of them, we use the experimental $f_{ij}$-values from \cite{ljung} and the data from the VALD database. For the remaining transitions, we apply $f_{ij}$-values calculated in this study.

{\it Photoionization cross sections}. Since there are no accurate data on the photoionization cross sections for Zr~I and Zr~II, we apply a hydrogenic approximation with an effective principal quantum number $n_{eff}$:
$$
n_{eff}=Z_{eff}\sqrt{\frac{\chi_H}{\chi}} ,
$$

\noindent where $Z_{eff}$ is the effective atomic charge ($Z_{eff}$ = 1 for neutral atoms, $Z_{eff}$ = 2 for the first ions, etc.),  $\chi_H$ is the ionization energy of the hydrogen atom from its ground state, and $\chi$ is the level ionization energy.

{\it Collisional rates}. The electron impact excitation was taken into account using the formula from \cite{vanreg} for permitted transitions. For forbidden transitions, the effective collision strength was assumed to be $\Upsilon$ = 1. The electron impact ionization was taken into account using the formula from \cite{drav}. In the atmosphere of cool stars, the number density of neutral hydrogen atoms is much greater than that for electrons. Therefore, it is also important to take into account the inelastic collisions with neutral hydrogen atoms. We used the Steenbock \& Holweger formula from \cite{sh} for permitted transitions and the approximation of Takeda \cite{takeda} for forbidden transitions. Since both formulas provide an accuracy only in order of magnitude, we introduced a scaling factor \kH\ in the range from 0 to 1, which we attempted to determine by analyzing the solar Zr~I and Zr~II lines.

\subsection{Mechanisms of departure from LTE}

In all our calculations, we used plane-parallel, homogeneous, blanketed model atmospheres computed using the MAFAGS code \cite{mafag}. The combined system of radiation transfer and statistical equilibrium equations was solved using the DETAIL code \cite{detail} based on the accelerated $\Lambda$ iteration method. The level populations obtained were used to calculate the profiles of the lines under study. Synthetic spectra were computed using the SIU code developed at the Munich University \cite{reetz}. The list of spectral lines includes all atomic and molecular lines from the tables by Kurucz \cite{kur}.

\begin{figure}

\resizebox{160mm}!{\includegraphics{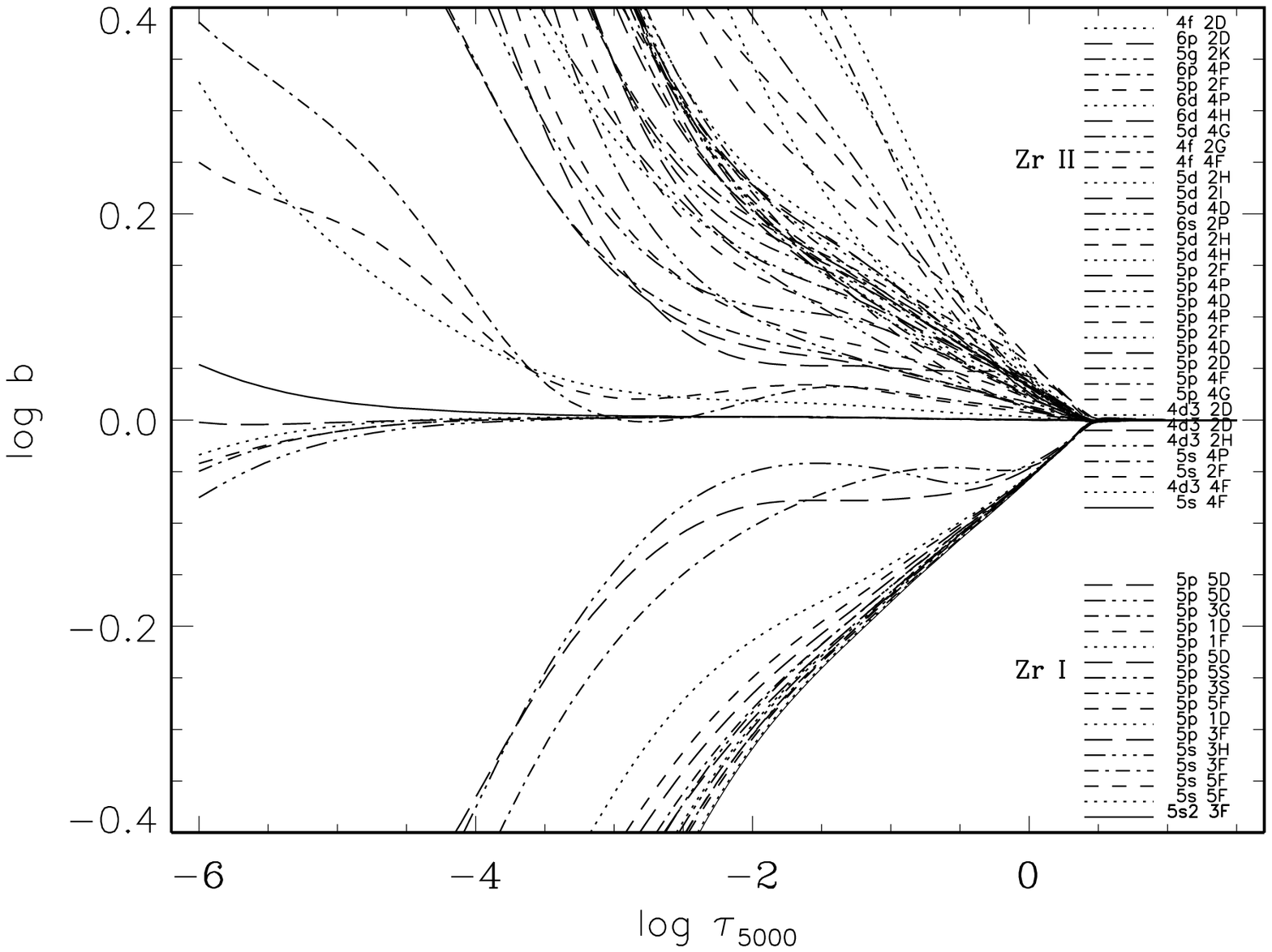}}
\caption{Departure coefficients $\log b$ for selected levels of Zr~I and Zr~II as a function of optical depth $\log \tau_{5000}$ in the solar model atmosphere. Departure coefficients of the ground states of Zr~I and Zr~II are plotted by continuous curves.}
\label{bfact}
\end{figure}

The departures from LTE in the level population are characterized by $b-$factor: $b_i = n_i/n_i*$, where $n_i$ is the level population obtained from the SE equations (non-LTE population) and $n_i*$ is the LTE population calculated from the Boltzmann-Saha formulas. 
Figure \ref{bfact} shows the $b-$factors of some Zr~I and Zr~II levels in the solar atmosphere. 
%Here, we investigate the pattern and causes of the departures of Zr~I and Zr~II level populations from their LTE values. 
The ground state and low excitation levels of Zr~II keep their TE populations almost over the entire atmosphere. Only in the uppermost layers with log$\tau_{5000} < -3$, the low excitation levels are slightly underpopulated, while the ground state is overpopulated. All of the remaining levels are overpopulated throughout the atmosphere, except for the 4d$^2$5p~$^4$F$^o_{3/2}$ level with \Eexc\,= 3.69~eV at depths around log$\tau_{5000}$ $= -3$. The overpopulation of high-excitation Zr~II levels results from radiative pumping of the ultra-violet (UV) transitions arising from the ground state and low-excitation levels. The main non-LTE mechanism for Zr~I is the overionization caused by superthermal UV radiation of a non-local origin below the thresholds of the low excitation levels. 
All Zr~I levels are underpopulated upwards log$\tau_{5000}$ $\approx$ 0.5.

The Zr~II lines used in the abundance determinations are formed in the solar atmosphere between log$\tau_{5000}$ = 0 and log$\tau_{5000} = -2$. In these layers, the upper level of each transition is overpopulated relative to its LTE population, while the populations of the lower levels keep their TE values. The line source function is defined as
$$
S_{ij}=\frac{2h\nu^3}{c^2}\frac{1}{\frac{b_i}{b_j}exp(h\nu/kT)-1},
$$
\noindent where $b_j$ and $b_i$ are the $b-$factors of the upper and lower levels, respectively. For the lines in the visual spectral range ($h\nu > kT$), the following approximate expression is valid:
$$
S_{ij}\simeq\frac{b_j}{b_i}B_{\nu}(T),
$$
\noindent where $B_{\nu}(T)$ is the Planck function.
In the line formation layers, $b_j > b_i$ is valid for each Zr~II line (Fig.\,\ref{bfact}), and the line source function is greater than the Planck function. As a result, the line is weakened compared to LTE and the zirconium abundance in non-LTE calculations should be increased to describe its LTE profile and equivalent width. Thus, the non-LTE effects lead to positive non-LTE corrections for Zr~II lines.

Non-LTE leads to a weakening of the Zr~I lines, too. However, the mechanism of departures from LTE is different from that for Zr~II lines. The Zr~I lines are weaker relative to their LTE 
strengths mainly due to the general overionization and also due
to $b_j/b_i > 1$ resulting in $S_{lu} > B_\nu$ and the depleted line absorption.At the depths where the Zr~I lines are formed, the upper levels are all
depleted to a lesser extent relative to their LTE populations than
are the lower levels. The non-LTE abundance corrections are positive for the Zr~I lines.

\section{Analysis of the solar zirconium lines}\label{Sun}

\subsection{Selection of lines}

Ljung~et~al. \cite{ljung} and Biemont and Grevesse \cite{biemont} provided the lists of lines which were used to determine the zirconium abundance of the solar atmosphere. We checked all lines from these lists for the presence of blends whose influence was difficult or impossible to take into account. For this purpose, we computed synthetic spectra using the SIU code by taking into account all of the atomic and molecular lines from the list by Kurucz \cite{kur} and compared them with the observed solar spectrum in fluxes \cite{sol-at} with a resolution R = 340 000 in the range 4000–4700~$\ang$ and R = 500 000 at longer wavelengths. We use a solar model atmosphere with T$_{eff}$ = 5780 K, logg = 4.44, and [M/H] = 0.0. The microturbulence velocity was assumed to be fixed, V$_{mic}$ = 0.9 km s$^{-1}$. To describe the observed profiles, the theoretical line profiles were convolved with a profile that combines
a rotational broadening of 1.8\,\kms\ and broadening by
macroturbulence with a radial-tangential profile of 3\,\kms\ to 4\,\kms\ for different lines.

%the rotation profile for velocity V$_{rot}sini$ = 1.8 km s$^{-1}$ and with the radial-tangential profile of macroturbulence motions with a velocity whose value was varied in the range 3-4 km s$^{-1}$ for different lines.

\begin{table}
\caption{Atomic data for the Zr~I lines, solar equivalent widths, and LTE abundances derived in this study and by Bi\'{e}mont \& Grevesse \cite{biemont}.}
\label{zr1lines}
\bigskip
\begin{tabular}{cccccl}
\hline
$\lambda^B$, $\ang$ $\rule{0pt}{15pt}$& $\chi_{exc}$, eV & log$gf^B$ & W$_{\lambda}^B$, m$\ang$ & log$\varepsilon_{Zr}$ & transition/comment\\
\hline
\hline
4241.706 & 0.65  &  0.14 &  3.7  & 2.33$^V$ $\rule{0pt}{15pt}$    & 4d$^3$5s $^5$F$_{3}$ - 4d$^3$5p $^5$F$^o_{3}$ \\
4687.805 & 0.73  &  0.55 &  10.0 & 2.34$^V$ & 4d$^3$5s $^5$F$_{5}$ - 4d$^3$5p $^5$G$^o_{6}$  \\
\multicolumn{6}{c}{ Lines excluded from zirconium abundance determination   }                     \\
3509.331 & 0.07  & -0.11 &  6.5  & 2.45$^B$ &   molecules                                        \\
3601.198 & 0.15  &  0.47 &  13   & 2.29$^B$ &   unknown absorbtion                           \\
3891.383 & 0.15  & -0.10 &  17   & 2.89$^B$ &   molecules                                         \\
4028.930 & 0.52  & -0.72 &  0.6  & 2.28$^B$ &   weak                                           \\
4030.049 & 0.60  & -0.36 &  2.6  & 2.65$^B$ &   weak + molecules                               \\
4043.608 & 0.52  & -0.37 &  5.8  & 2.95$^B$ & Nd II $4043.595\ang$                   \\
4072.695 & 0.69  &  0.31 &  5.7  & 2.42$^B$ &   molecules                                         \\
4507.100 & 0.54  & -0.43 &  3.6  & 2.76$^B$ &   molecules + atoms                                \\
4542.234 & 0.63  & -0.31 &  4.6  & 2.83$^B$ &   weak + molecules                                 \\
4710.077 & 0.69  &  0.37 &  10.5 & 2.59$^B$ &    molecules                                       \\
4732.323 & 0.63  & -0.49 &  2.5  & 2.72$^B$ &    weak + molecules                               \\
4739.454 & 0.65  &  0.23 &  5.5  & 2.38$^B$ &   unknown absorbtion                           \\
4772.310 & 0.62  &  0.04 &  5.3  & 2.52$^B$ &   unknown absorbtion                           \\
4784.940 & 0.69  & -0.49 &  1.6  & 2.57$^B$ &    weak + molecules                              \\
4805.890 & 0.69  & -0.42 &  1.5  & 2.47$^B$ &    weak + molecules                                \\
4809.477 & 1.58  &  0.16 &  1.6  & 2.77$^B$ &  weak + molecules                                \\
4815.056 & 0.65  & -0.53 &  2.0  & 2.67$^B$ &    weak + molecules                               \\
4815.637 & 0.60  & -0.03 &  3.0  & 2.30$^B$ &    weak + molecules                              \\
4828.060 & 0.62  & -0.64 &  1.9  & 2.73$^B$ &  weak + molecules                               \\
5046.550 & 1.53  &  0.06 &  0.50 & 2.29$^B$ &   weak + molecules                               \\
5385.128 & 0.52  & -0.71 &  1.8  & 2.63$^B$ &    weak + molecules                               \\
6127.460 & 0.15  & -1.06 &  2.1  & 2.63$^B$ &    weak + molecules                               \\
6134.570 & 0.00  & -1.28 &  1.9  & 2.66$^B$ &    weak + molecules                               \\
6140.460 & 0.52  & -1.41 &  0.73 & 2.88$^B$ &     weak + molecules                               \\
6143.183 & 0.07  & -1.10 &  2.1  & 2.59$^B$ &   weak + molecules                               \\
6313.030 & 1.58  &  0.27 &  1.1  & 2.40$^B$ &   weak + molecules                               \\
6445.720 & 1.00  & -0.83 &  0.94 & 2.86$^B$ &    weak + molecules                               \\
6990.840 & 0.62  & -1.22 &  0.50 & 2.57$^B$ &     weak + molecules                               \\
7097.760 & 0.69  & -0.57 &  2.1  & 2.61$^B$ & weak + molecules                               \\
7102.890 & 0.65  & -0.84 &  0.65 & 2.33$^B$ &    weak + molecules                               \\
7819.350 & 1.82  & -0.38 &  0.65 & 2.96$^B$ &    weak + molecules                               \\
7849.380 & 0.69  & -1.30 &  1.0  & 2.97$^B$ &   weak + molecules                               \\
\hline
\multicolumn{6}{l}{\small $^B$  - based on the data from \cite{biemont}  $\rule{0pt}{15pt}$  }                    \\
\multicolumn{6}{l}{\small $^V$  - this paper}                                                        \\
\end{tabular}
\end{table}

Tables \ref{zr1lines} and \ref{zr2lines} list all of the checked Zr~I and Zr~II lines. The oscillator strengths of Zr~II lines were taken from \cite{ljung} as the most recent ones and from \cite{biemont}. The data for Zr~I lines are available only in \cite{biemont}. All $f_{ij}$ were obtained from laboratory measurements. Below, we give a brief description of all lines.

\begin{figure}[h]

\includegraphics{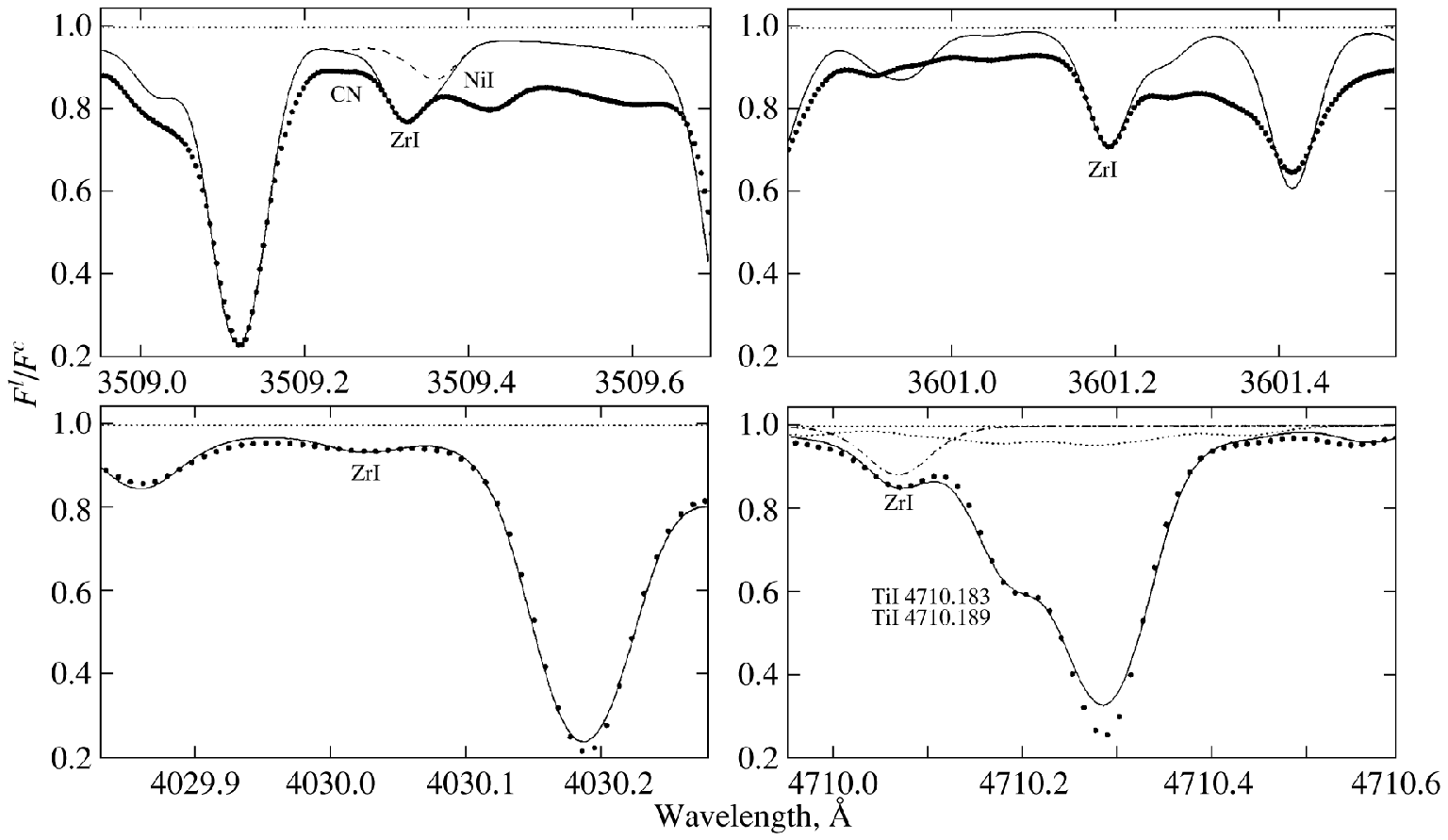}

\caption{Solar Zr~I lines excluded from the abundance determination. Synthetic LTE (continuous curve) flux profiles of the 3509, 3601, 4030, and 4710~$\ang$ blends compared with the observed spectrum of the Kurucz et~al. \cite{sol-at} solar flux atlas (bold dots). Dash-dotted curve shows the pure Zr~I 4710~$\ang$ line, while dotted curve shows the molecular C$_2$ lines. Dashed curve plots the synthetic spectrum of the 3509~$\ang$ blend calculated with no zirconium in the atmosphere.}
\label{zr1exc}
\end{figure}

{\it Zr~I lines}. Biemont and Grevesse \cite{biemont} used 34 Zr~I lines to determine the zirconium abundance. It emerged that almost all of the lines from this list, except two lines, Zr~I 4241.706 and 4687.799~$\ang$, were strongly blended and we excluded them from the abundance determinations. Let us illustrate this with several examples. The strongest Zr~I 3891.383~$\ang$ line lies in the wing of the H$_{\zeta}$ 3889.051~$\ang$ line and is described without confidence. The Zr~I lines are mostly very weak - only four of them have an equivalent width larger than 10~m$\ang$. Therefore, they can be affected even by  weak lines of other atoms and molecules. Figure~\ref{zr1exc} shows our attempts to describe the blends near four Zr~I lines. Zr~I 4710.077~$\ang$ is one of the strongest lines, but it is located in a molecular band. Besides, it is influenced by two strong lines, Ti~I 4710.183~$\ang$ and Ti~I 4710.189~$\ang$. The Zr~I 3509.331~$\ang$ line is influenced by the CN 3509.247~$\ang$ and Ni~I 3509.362~$\ang$ lines and by the absorption of an unknown origin in the range 3509.2-3509.6~$\ang$. 
We are unable to properly take into account the absorption around the Zr~I 3601.198~$\ang$ line, because we do not know what is responsible for it. The Zr~I 4030.049~$\ang$ line is weak (W$_{\lambda}$ = 2.6~m$\ang$, Table~\ref{zr1lines}) and is lost in the absorption produced by the wings of the strong Fe~I 4030.185~$\ang$ line.

The profiles of the strong Zr~I 4043, 4072, 4739, 4772~$\ang$ lines are also distorted by the influence of molecular and atomic lines. In addition, in most cases, there is absorption near the zirconium lines that we are unable to take into account, because we do not know what is responsible for it. The profiles of the lines being studied are generally asymmetric, suggesting the presence of other absorption sources at close wavelengths.

\begin{figure}
\includegraphics{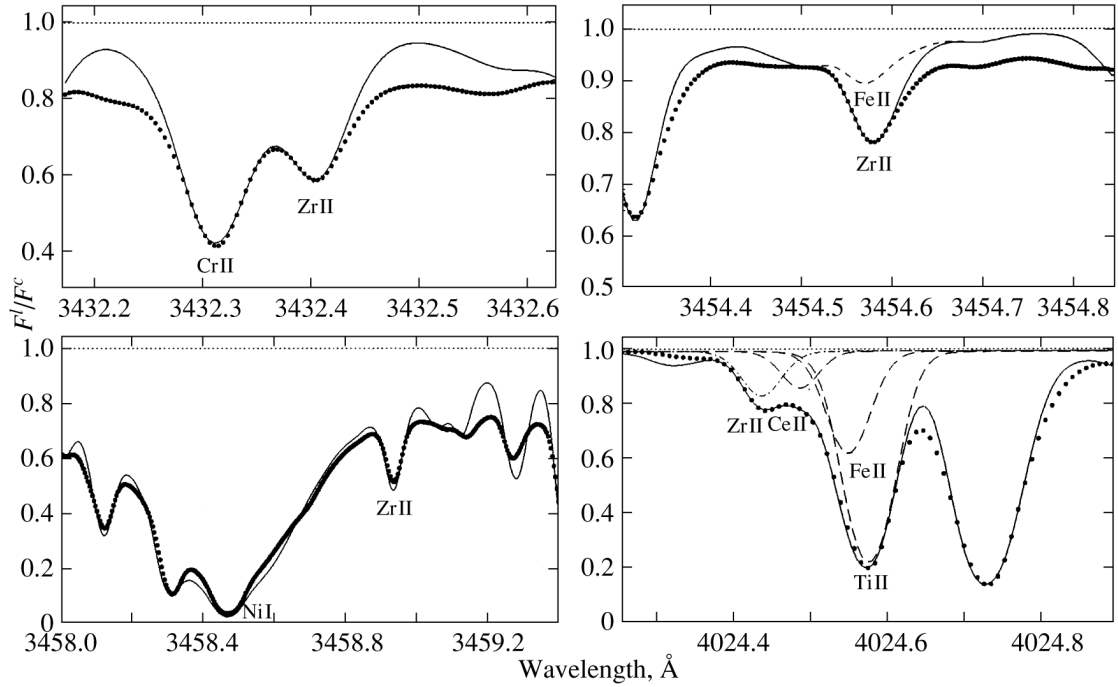}
\caption{Solar lines of Zr~II excluded from the abundance determination. Synthetic LTE (continuous curve) flux profiles of the 3432, 3454, 3458, and 4024~$\ang$ blends compared with the observed spectrum of the Kurucz et~al. \cite{sol-at} solar flux atlas (bold dots). Dash-dotted curve shows the pure Zr~II 4024~$\ang$ line, while long-dashed curve shows the lines of Ce~II, Fe~II, and Ti~I. Dashed curve plots the synthetic spectrum of the 3454~$\ang$ blend calculated with no zirconium in the atmosphere.}
\label{zr2exc}
\end{figure}

{\it Zr~II lines}. We checked 25 Zr~II lines, 16 of which turned out to be unsuitable for the abundance determination. Let us illustrate this with examples. Figure~\ref{zr2exc} shows examples of the Zr~II lines excluded from the subsequent analysis. The Zr~II 3458.932~$\ang$ is located in the wing of the strong Ni~I 3458.456~$\ang$ line and is also blended with Ti~II 3458.903~$\ang$. As can be seen from Fig.~\ref{zr2exc}, we cannot describe well the local continuum near the Zr~II 3458.932~$\ang$ line. The Zr~II 3454.572~$\ang$ line is blended with Fe~I 3454.569~$\ang$ and we cannot reproduce the local continuum. We cannot describe the absorption near Zr~II 3432.404~$\ang$ either, because we do not know was is responsible for it.

The Zr~II 4024.435~$\ang$ line from the list by Ljung et~al. \cite{ljung} deserves particular attention. Despite good agreement with the abundances determined by Ljung et~al. \cite{ljung} from different lines, we believe that it cannot be used in our analysis because of strong blending. The Zr~II 4024.435~$\ang$ line is influenced by three lines at once: Ce~II 4024.485~$\ang$, Fe~II 4024.547~$\ang$, and Ti~I 4024.572~$\ang$ (Fig.~\ref{zr2exc}). To achieve the best agreement between the observed and theoretical spectra, the oscillator strengths of the cerium and titanium lines were reduced by 0.07 and 0.10~dex, respectively. The figure shows the contribution from each of the lines to the total absorption. We exclude the Zr~II 4024.435~$\ang$ line from the subsequent analysis.

The Zr~II 3549, 3588~$\ang$ lines are located in molecular bands. The Zr~II 3479.017, 3499, 3607, 3671, 3714, 3796, 3836, 4034, 4085, 4317~$\ang$ lines are strongly influenced by the lines of other atoms and the accuracy of determining the zirconium abundance depends strongly on the accuracy of the atomic parameters of the blending lines.
Below, we give the list of Zr~I and Zr~II lines used to determine the zirconium abundance with the corresponding comments.

\begin{figure}
\includegraphics{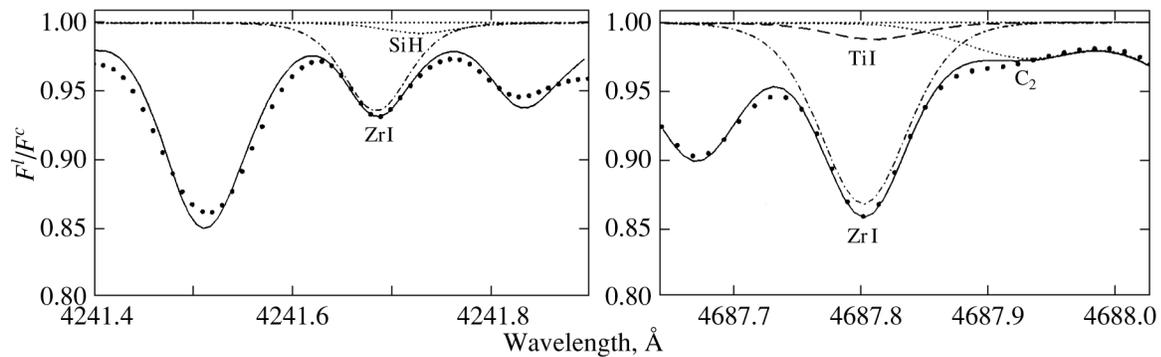}
\caption{Synthetic LTE (continuous curve) flux profiles of the selected Zr~I lines compared with the observed spectrum of the Kurucz et~al. \cite{sol-at} solar flux atlas (bold dots). Dash-dotted curve shows the pure Zr~I 4241 and 4687~$\ang$ lines. Dotted curve shows the molecular SiH and C$_2$ lines. Long-dashed curve shows the Ti~I line.}
\label{zr1inc}
\end{figure}

{\it Zr~I 4241.706~$\ang$}. As can be seen from Fig.~\ref{zr1inc}, the Zr~I 4241.706~$\ang$ line is influenced by the SiH molecular line (W$_\lambda$ = 0.6 m$\ang$, 11 of the total equivalent width). As $f_{ij}$(SiH) increases by 0.2~dex, the abundance determined from the Zr~I 4241.706~$\ang$ line decreases by 0.02~dex. The total profile is described more poorly.

{\it Zr~I 4687.805~$\ang$}. This line is blended with the Ti~I 4687.809~$\ang$ line (W$_\lambda$ = 1.1 m$\ang$) and the right wing of the Zr~I 4687.805~$\ang$ line being studied is influenced by the C$_2$ 4687.9~$\ang$ molecular line, as shown in Fig.~\ref{zr1inc}. Changing $f_{ij}$(C$_2$) by 0.5~dex does not lead to any change in the zirconium abundance determined from the Zr~I 4687.805~$\ang$ line. The atomic data for Ti~I 4687.809~$\ang$ were taken from VALD.

\begin{figure}
\includegraphics{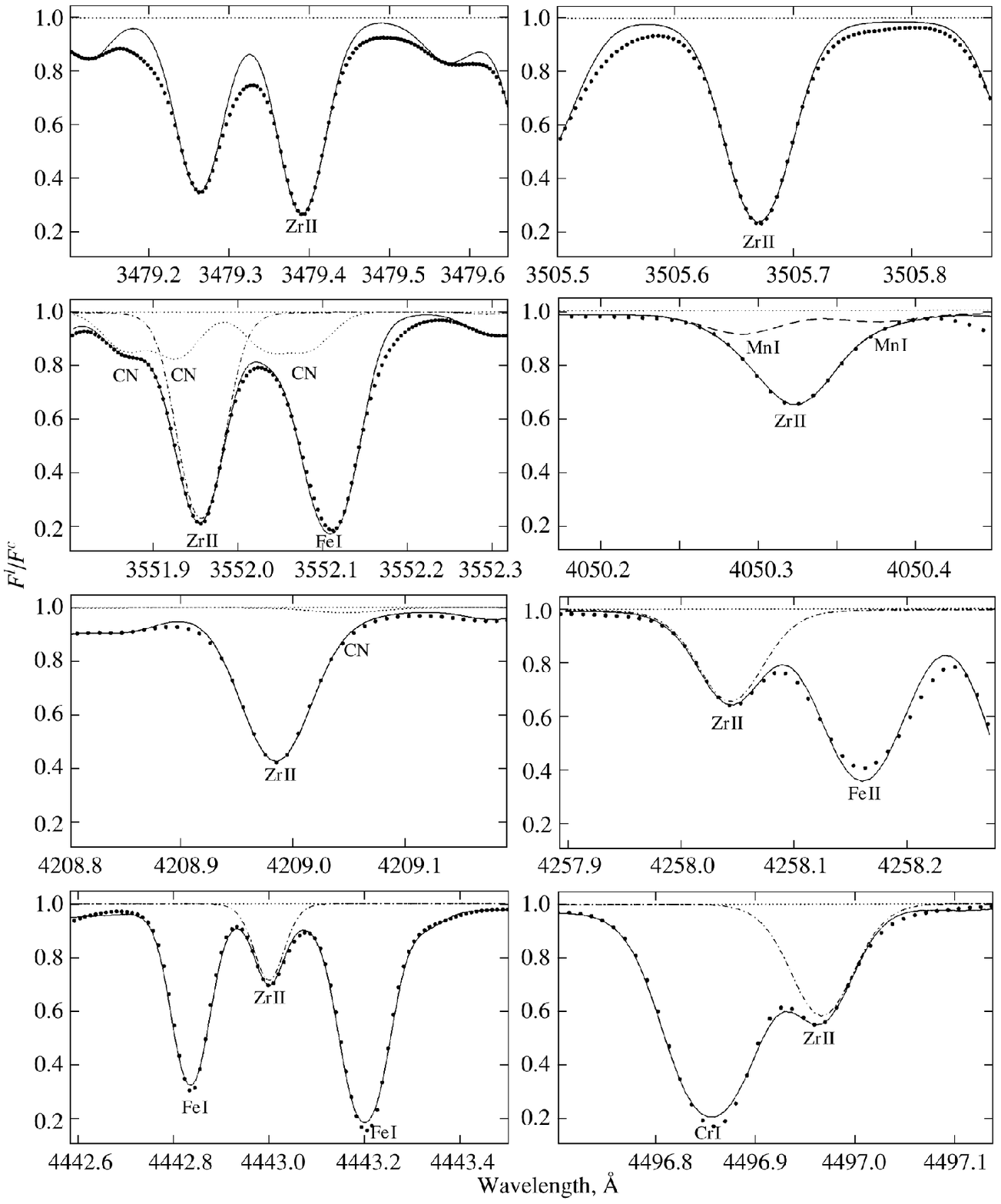}
\caption{Synthetic LTE (continuous curve) flux profiles of the selected Zr~II lines compared with the observed spectrum of the Kurucz et~al. \cite{sol-at} solar flux atlas (bold dots). Dash-dotted curve shows the pure Zr~II 4258, 4443, and 4496~$\ang$ lines. Dotted curve shows the molecular CN lines. Long-dashed curve shows the Mn~I lines.}
\label{zr2inc}
\end{figure}

{\it Zr~II 3479.387~$\ang$}. As can be seen from Fig.~\ref{zr2inc}, the observed wings of this line lie below the theoretical ones approximately by 5\%, suggesting that we do not completely take into account the absorption in the local continuum. When the local continuum was raised by 5\%, the equivalent width of the Zr~II 3479.387~$\ang$ line decreased by 6.4\%. In this case, the zirconium abundance decreased by 0.13~dex.

{\it Zr~II 3505.666~$\ang$}. The theoretical level of the local continuum near this line is above the observed one by 2\%. Since the line is strong (W$_\lambda$ = 53.4 m$\ang$), this affects insignificantly the abundance determination from this line.When the local continuum is raised by 2\%, the line equivalent width W$_\lambda$(Zr~II) decreases by 3.7\% and the abundance decreases by 0.07~dex.

{\it Zr~II 3551.951~$\ang$}. This line is located in the CN molecular band whose influence is weak (Fig.~\ref{zr2inc}).

{\it Zr~II 4050.320~$\ang$}. This line is blended with two lines of manganese, Mn~I 4050.288 and 4050.345~$\ang$ (Fig.~\ref{zr2inc}), whose abundance in the solar atmosphere is known. Therefore, we use the Zr~II 4050.320~$\ang$ line in analyzing the zirconium abundance only in the solar atmosphere. However, it should be used with caution for other stars and only when the manganese abundance is determined carefully.

{\it Zr~II 4208.980~$\ang$}. The red wing of this line is slightly influenced by the CN 4209.046~$\ang$ molecular line. According to the list by Kurucz \cite{kur}, the equivalent width of the molecular line should not exceed 1.4 m$\ang$, while W$_\lambda$(Zr~II 4208~$\ang$) = 44.4 m$\ang$. Figure~\ref{zr2inc} describes the observed spectrum near the Zr~II
4208.980~$\ang$ line with allowance made for all lines and separately shows the contribution from the molecular line. Increasing log$gf$(CN) by 0.1~dex causes the zirconium abundance determined from the Zr~II 4208.980~$\ang$ line to decrease by less than 0.01~dex.
\begin{table}

\caption{Atomic data for the Zr~II lines, solar equivalent widths, and LTE abundances derived in this study, Biemont \& Grevesse \cite{biemont}, and Ljung et~al. \cite{ljung}.}
\label{zr2lines}
\bigskip
\begin{tabular}{cccccl}
\hline
$\lambda^L$, $\ang$ $\rule{0pt}{15pt}$& ~~$\chi_{exc}$, eV ~~& ~~log$gf^L$ ~~& ~~W$_{\lambda}$, m$\ang$~~ & ~~log$\varepsilon_{Zr}~~$ & comment\\
\hline
\hline
3479.387 & 0.713 & -0.18 &  55.7$^V$ & 2.66$^V$ & 4d$^2$5s $^2$F$_{5/2}$ - 4d$^2$5p$^2$G$^o_{7/2}$  \\
3505.666 & 0.164 & -0.39 &  53.4$^V$ & 2.63$^V$ & 4d$^2$5s $^4$F$_{9/2}$ - 4d$^2$5p $^4$G$^o_{9/2}$ \\
3551.951 & 0.095 & -0.36 &  53.9$^V$ & 2.49$^V$ & 4d$^2$5s $^4$F$_{7/2}$ - 4d$^2$5p $^4$G$^o_{7/2}$ \\
4050.320 & 0.713 & -1.06 &  18.6$^V$ & 2.42$^V$ & 4d$^2$5s $^2$F$_{5/2}$ -  4d$^2$5p $^2$D$^o_{5/2}$ \\
4208.980 & 0.713 & -0.51 &  44.4$^V$ & 2.65$^V$ & 4d$^2$5s $^2$F$_{5/2}$ - 4d$^2$5p $^2$F$^o_{5/2}$  \\
4258.041 & 0.559 & -1.20 &  25.7$^V$ & 2.61$^V$ & 4d$^2$5s $^2$D$_{5/2}$ - 4d$^2$5p $^4$G$^o_{5/2}$  \\
4442.992 & 1.486 & -0.42 &  24.1$^V$ & 2.65$^V$ & 4d$^3$ $^2$H$_{7/2}$ - 4d$^2$5p $^2$G$^o_{7/2}$    \\
4496.962 & 0.713 & -0.89 &  34.5$^V$ & 2.67$^V$ & 4d$^2$5s $^2$F$_{5/2}$ - 4d$^2$5p $^4$G$^o_{5/2}$  \\
5112.270 & 1.665 & -0.85 &  9.3$^V$  & 2.67$^V$ & 4d5s$^2$ $^2$D$_{3/2}$ - 4d$^2$5p $^2$D$^o_{3/2}$  \\
\multicolumn{6}{c}{Lines excluded from zirconium abundance determination }                           \\
3432.404 & 0.931 & -0.72 &  21$^B$   & 2.69$^B$ &                                                    \\
3454.572 & 0.931 & -1.33 &  10.0$^B$ & 2.83$^B$ &                                                    \\
3458.932 & 0.959 & -0.48 &  16$^B$   & 2.31$^B$ &                                                    \\
3479.017 & 0.527 & -0.67 &  28$^B$   & 2.47$^B$ &                                                    \\
3499.571 & 0.409 & -1.06 &  24$^B$   & 2.33$^B$ &                                                    \\
3549.508 & 1.236 & -0.72 &  16$^B$   & 2.43$^B$ &                                                    \\
3588.314 & 0.409 & -1.13 &  27$^B$   & 2.73$^B$ &                                                    \\
3607.369 & 1.236 & -0.70 &  13$^B$   & 2.54$^B$ &                                                    \\
3671.264 & 0.713 & -0.58 &  32$^B$   & 2.53$^B$ &                                                    \\
3714.777 & 0.527 & -0.96 &  30$^B$   & 2.64$^B$ &                                                    \\
3796.482 & 1.011 & -0.89 &  15$^B$   & 2.48$^B$ &                                                    \\
3836.761 & 0.559 & -0.12 &  47$^B$   & 2.28$^B$ &                                                    \\
4024.435 & 0.999 & -1.13 &  12.0$^L$ & 2.56$^L$ &                                                    \\
4034.083 & 0.802 & -1.51 &  6.5$^B$  & 2.53$^B$ &                                                    \\
4085.719 & 0.931 & -1.84 &  5.4$^B$  & 2.62$^B$ &                                                    \\
4317.309 & 0.713 & -1.45 &  12.0$^B$ & 2.55$^B$ &                                                    \\
\hline
\multicolumn{6}{l}{\small $^B$ - based on data from Bi\'{e}mont \& Grevesse \cite{biemont} $\rule{0pt}{15pt}$}  \\
\multicolumn{6}{l}{\small $^L$ - based on data from Ljung et al \cite{ljung}   }                        \\
\multicolumn{6}{l}{\small $^V$ - this work }                        \\
\end{tabular}
\end{table}

{\it Zr~II 4258.041~$\ang$}. The stronger Fe~II 4258.154~$\ang$ line affects only the red wing of the zirconium line (Fig.~\ref{zr2inc}). To describe well the iron line, we had to increase its wavelength by 0.06~$\ang$ and the oscillator strength by 0.28~dex. However, this increase in log$gf$(Fe~II) had virtually no effect on the abundance determined from the Zr~II 4258.041~$\ang$ line ($\Delta(\rm log \it \varepsilon_{Zr}) <$ 0.01~dex).

{\it Zr~II 4442.992~$\ang$}. The zirconium line lies between two strong lines, Fe~I 4442.832 and 4443.194~$\ang$ (Fig.~\ref{zr2inc}). For the best description of the observed spectrum, we had to increase the oscillator strength of the Fe~I 4442.832~$\ang$ line by 0.05~dex. Increasing the oscillator strengths of both iron lines by 0.05~dex causes the zirconium abundance determined from the Zr~II 4442.992~$\ang$ line to decrease by 0.01~dex.

{\it Zr~II 4496.962~$\ang$}. This line is located in the wing of the Cr~I 4496.854~$\ang$ line (Fig.~\ref{zr2inc}). We determined the zirconium abundance from the red wing of the Zr~II 4496.962~$\ang$ line. The oscillator strength had to be increased by 0.1~dex, which had no effect on the zirconium abundance determination.

{\it Zr~II 5112.270~$\ang$}. This line is virtually isolated. The C$_2$ 5112.361~$\ang$ molecular line makes a small contribution to the absorption. Changing log$gf$(C$_2$) by 0.1 does not lead to any change in the zirconium abundance determined from the Zr~II 5112.270~$\ang$ line.

We concluded that most of the lines from the list by Biemont and Grevesse \cite{biemont} and one Zr~II line from the list by Ljung et~al. \cite{ljung} cannot be used to determine the zirconium abundance, because the contribution from neighboring lines cannot be properly taken into account, especially when the absorption near the line under study is produced by unknown sources (i.e., the lines are absent in the list by Kurucz \cite{kur}). Only two Zr~I lines and nine Zr~II line can be used to reliably estimate the zirconium abundance in the solar atmosphere.

Zirconium in nature is represented by five isotopes: $^{90}$Zr, $^{91}$Zr, $^{92}$Zr, $^{94}$Zr, and $^{96}$Zr. In the matter of the Solar system, the relative contribution from each of the isotopes to the total zirconium abundance is 51.45:11.22:17.15:17.38:2.80 \cite{lodders}. None of the works on the experimental study of the of the zirconium spectrum provides data on the isotopic shifts and hyperfine level splitting. Therefore, we considered the zirconium lines to be one-component ones.

\subsection{Solar zirconium abundance}

Using the selected Zr~I and Zr~II lines, we determined the mean zirconium LTE abundances for each ionization stage (Table~\ref{nlte-tab}). The difference $\log\varepsilon$(Zr~II) - $\log\varepsilon$(Zr~I) = 0.28~dex turned out to exceed considerably the error in the abundance determined from individual lines.

\begin{table}[h!]

\caption{The solar Zr abundances from the individual Zr~I and Zr~II lines and the mean values derived from the LTE and NLTE ($k_H$ = 0, 0.1, 0.33, 1, 0.1+, 0.33+, and 1+) calculations with the MAFAGS solar model atmosphere of T$_{\rm eff}$ = 5780~K, $\log g$ = 4.44, $V_{mic}$ = 0.9 km/s.}
\label{nlte-tab}
\bigskip
\begin{tabular}{p{1.7cm}|p{0.5cm}p{0.5cm}p{0.7cm}c|*{9}{p{0.5cm}}p{0.7cm}c}
%\begin{tabular}{l|c|c|c|c|c|c|c|c|c|c|c|c|c|c|c}
\hline
version   & \multicolumn{4}{c|} {Zr~I} & \multicolumn{11}{c}{Zr~II} \\
&  4241 & 4687 & log$\varepsilon_{Zr}$ & $\sigma$ & 3479 & 3505 & 3551 & 4050 & 4208 & 4258 & 4442 & 4496 & 5112 & log$\varepsilon_{Zr}$ & $\sigma$ \\
\hline
\hline
LTE        & 2.33 & 2.34 & 2.33 & 0.01 & 2.66 & 2.63 & 2.49 & 2.42 & 2.65 & 2.61 & 2.65 & 2.67 & 2.67 & 2.61 & 0.09  \\
NLTE       &      &      &      &      &      &      &      &      &      &      &      &      &      &      &       \\
$k_H$=0.00 & 2.68 & 2.61 & 2.65 & 0.05 & 2.77 & 2.67 & 2.54 & 2.50 & 2.76 & 2.65 & 2.73 & 2.75 & 2.82 & 2.69 & 0.11  \\
$k_H$=0.10 & 2.68 & 2.61 & 2.65 & 0.05 & 2.77 & 2.67 & 2.54 & 2.50 & 2.76 & 2.65 & 2.72 & 2.74 & 2.81 & 2.68 & 0.11  \\
$k_H$=0.33 & 2.67 & 2.59 & 2.63 & 0.06 & 2.76 & 2.66 & 2.53 & 2.49 & 2.75 & 2.64 & 2.72 & 2.73 & 2.80 & 2.68 & 0.11  \\
$k_H$=1.00 & 2.65 & 2.57 & 2.61 & 0.06 & 2.76 & 2.65 & 2.52 & 2.48 & 2.73 & 2.64 & 2.71 & 2.72 & 2.79 & 2.67 & 0.11  \\
$k_H$=0.10+& 2.62 & 2.61 & 2.62 & 0.01 & 2.74 & 2.65 & 2.52 & 2.45 & 2.70 & 2.62 & 2.70 & 2.70 & 2.72 & 2.64 & 0.10  \\
$k_H$=0.33+& 2.57 & 2.58 & 2.58 & 0.01 & 2.72 & 2.64 & 2.51 & 2.43 & 2.68 & 2.62 & 2.67 & 2.69 & 2.69 & 2.63 & 0.10  \\
$k_H$=1.00+& 2.53 & 2.54 & 2.54 & 0.01 & 2.70 & 2.63 & 2.50 & 2.42 & 2.66 & 2.61 & 2.66 & 2.68 & 2.68 & 2.62 & 0.09  \\
\hline
\end{tabular}
\end{table}

When the LTE assumption is abandoned, the accuracy of the derived zirconium abundance depends on the accuracy of the photoionization cross sections $\sigma_{ph}$, transition probabilities, and the rates of inelastic collisions with electrons, $C_{ij}(e)$, and neutral hydrogen atoms, $C_{ij}(H)$. The influence of an uncertainty in $\sigma_{ph}$ and $C_{ij}(e)$ on the results will be considered in the next subsection. Here, we will talk about the collisions with hydrogen atoms and about the oscillator strengths.

\begin{figure}
\includegraphics{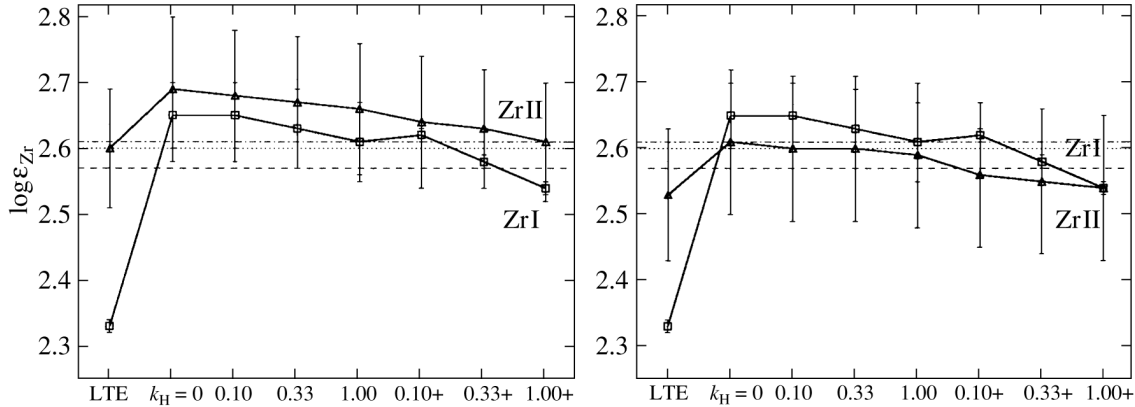}
\caption{The mean solar Zr abundances derived from the lines of Zr~I (squares) and Zr~II (triangles) using the LTE and NLTE ($k_H$ = 0, 0.1, 0.33, 1, 0.1+, 0.33+, and 1+) calculations. In left and right panel, we use oscillator strengths of the Zr~II lines from Ljung et~al. \cite{ljung} and Bi\'{e}mont \& Grevesse \cite{biemont}, respectively. The meteoritic Zr abundance from Anders \& Grevesse \cite{anders}, Lodders et~al. \cite{lodders}, and Asplund et~al. \cite{asplund} is indicated by dash-dotted, dotted, and dashed curves, respectively.}
\label{nlte-ab}
\end{figure}

According to recent studies by Belyaev et~al. \cite{belyaev} and Belyaev and Barklem \cite{bel-bark} for the resonance Na~I and Li~I lines, the formula from Steenbock and Holweger \cite{sh} gives overestimated rates of collisions with hydrogen atoms. Therefore, we performed our calculations with a scale factor $k_H$ that was varied in the range from 0 to 1. We made an attempt to estimate $k_H$ empirically by analyzing Zr~I and Zr~II lines in the solar spectrum. The calculations were made for eight spectral line formation scenarios: LTE, non-LTE without allowance for the collisions with hydrogen atoms, non-LTE with allowance for the collisions with hydrogen atoms only in permitted transitions with $k_H$ = 0.1, 0.33, 1.0, non-LTE with allowance for the hydrogen collisions in both permitted and forbidden transitions with $k_H$ = 0.1, 0.33, 1.0 (below, we will denote them by 0.1+, 0.33+, and 1.0+). For each of the scenarios, we determined the mean abundances from the lines of two ionization stages, Zr~I and Zr~II. The results are presented in Fig.~\ref{nlte-ab}.

The left and right panels of Fig.~\ref{nlte-ab} differ only by the use of different sources of oscillator strengths for Zr~II lines, i.e., \cite{ljung} and \cite{biemont}, respectively. On the left panel, there exists a systematic abundance difference between the two ionization stages: the zirconium abundances derived from Zr~II lines are higher than those inferred from Zr~I lines independent of the line formation assumptions used. It should be noted that the oscillator strengths of Zr~II lines from Ljung et~al. \cite{ljung} are systematically lower than those from and Grevesse \cite{biemont} by a value from 0.03~dex for the Zr~II 3505~$\ang$ line to 0.26~dex for the Zr~II 5112~$\ang$ line. When the oscillator strengths from \cite{biemont} are used, the mean abundances derived from Zr~II lines decrease by 0.09~dex in all our non-LTE calculations and the systematic difference log$\varepsilon_{Zr}$(Zr ~II) — log$\varepsilon_{Zr}$(Zr~I) changes its sign (Fig.~\ref{nlte-ab}) compared to the case where $f_{ij}$ from Ljung et~al.\cite{ljung} are used for Zr~II lines. This points to the need for revising the values of $f_{ij}$ for Zr~I lines. Below, we consider only the case where $f_{ij}$ from Ljung et~al.\cite{ljung} are used for Zr~II lines. We see that any of the non-LTE scenarios leads to better agreement from the lines of the two ionization stages than the LTE case: the abundances inferred from Zr~I and Zr~II agree between themselves and with the meteorite abundances within 1$\sigma$. Since we cannot give preference to any of the non-LTE calculations, we abandon the attempt to empirically estimate $k_H$. Nevertheless, for the subsequent study, we dwelt on $k_H$ = 0.1+, for which the difference log$\varepsilon_{Zr}$(Zr~II) — log$\varepsilon_{Zr}$(Zr~I) is minimal and equal to 0.02~dex (see Table~\ref{nlte-tab}).

\subsection{Influence of the uncertainties in atomic parameters on the non-LTE results}

We performed test calculations by varying the photoionization cross sections $\sigma_{ph}$ and the rates of collisions with electrons $C_{ij}(e)$. We traced the changes in the abundance derived from all Zr~I and Zr~II lines when the photoionization cross sections increased and decreased by a factor of 10: $\sigma_{ph}(n_{eff})\times$10 and $\sigma_{ph}(n_{eff})$/10 (for the  standard calculations with $k_H$ = 0.1+ and using $n_{eff}$ in the formula for $\sigma_{ph}$). Similar calculations were made for $k_H$ = 0.1+ but using the principal quantum number n, i.e., with $\sigma_{ph}(n)$, $\sigma_{ph}(n)\times$10, $\sigma_{ph}(n)$/10. The results are presented in Table~\ref{delta-nltecorr}.

\begin{table}
\caption{Influence of an uncertainty in atomic parameters on the non-LTE zirconium abundance determination.}
\label{delta-nltecorr}
\bigskip
%\begin{tabular}{lccccccccccc}
%\begin{tabular}{p{4.5cm}|*{2}{p{0.8cm}}|*{9}{p{0.8cm}}}
\begin{tabular}{p{4.5cm}@{}|p{1.1cm}@{}p{1.1cm}@{}|p{1.1cm}@{}p{1.1cm}@{}p{1.1cm}@{}p{1.1cm}@{}p{1.1cm}@{}p{1.1cm}@{}p{1.1cm}@{}p{1.1cm}@{}p{1.1cm}@{}p{1.1cm}}
\hline
                                 &\multicolumn{11}{c}{Change in abundance (dex)}   $\rule{0pt}{12pt}$       \\
\cline{2-12}
Рarameter                        &  \multicolumn{2}{c|}{Zr~I} &\multicolumn{9}{c}{Zr~II}$\rule{0pt}{12pt}$\\
%\cline{2-12}
                                 & 4241 & 4687 & 3479 & 3505 & 3551 & 4050 & 4208 & 4258 & 4442 & 4496 & 5112$\rule{0pt}{12pt}$\\
\hline
\hline
Photoionization                    &\multicolumn{11}{c}{relative to log$\varepsilon_{Zr}(k_H = 0.1+,~  \sigma_{ph}(n_{eff}))$}$\rule{0pt}{12pt}$\\
$\sigma_{ph}(n_{eff})\times$10   & ~0.00 & ~0.00 & ~0.00 & ~0.00 & ~0.00 & ~0.00 & ~0.00 & ~0.00 & ~0.00 & ~0.00 & ~0.00 \\
$\sigma_{ph}(n_{eff})$/10        & -0.01 & -0.01 & ~0.00 & ~0.00 & ~0.00 & ~0.00 & ~0.00 & ~0.00 & ~0.00 & ~0.00 & ~0.00 \\
$\sigma_{ph}(n)$                 & -0.04 & -0.04 & ~0.00 & ~0.00 & ~0.00 & ~0.00 & ~0.00 & ~0.00 & ~0.00 & ~0.00 & ~0.00 \\
$\sigma_{ph}(n)\times$10         & ~0.01 & ~0.01 & ~0.00 & ~0.00 & ~0.00 & ~0.00 & ~0.00 & ~0.00 & ~0.00 & ~0.00 & ~0.00 \\
$\sigma_{ph}(n)$/10              & -0.09 & -0.09 & ~0.00 & ~0.00 & ~0.00 & ~0.00 & ~0.00 & ~0.00 & ~0.00 & ~0.00 & ~0.00 \\
Collisions with electrons                     &\multicolumn{11}{c}{relative to log$\varepsilon_{Zr}(k_H = 0.1+,~ C_{ij}(e))$}$\rule{0pt}{12pt}$\\
$C_{ij}(e)\times10$ ($k_H$=0.1+) & -0.03 & -0.01 & -0.02 & -0.01 & ~0.00 & -0.01 & ~0.00 & ~0.00 & -0.02 & -0.01 & -0.02 \\
                                 &\multicolumn{11}{c}{relative to log$\varepsilon_{Zr}(k_H = 0.0,~C_{ij}(e))$}$\rule{0pt}{12pt}$\\
$C_{ij}(e)\times10$ ($k_H$=0.0)  & -0.05 & -0.01 & -0.03 & -0.02 & -0.02 & -0.05 & -0.04 & -0.02 & -0.04 & -0.05 & -0.08 \\

\hline
\end{tabular}
\end{table}

When the principal quantum number is used instead of $n_{eff}$, the non-LTE effects for Zr~I lines are reduced: the correction to the abundance for both Zr~I lines is 0.04~dex. This is because $n$ > $n_{eff}$ for all Zr~I levels; therefore, $\sigma_{ph}(n)$ < $\sigma_{ph}(n_{eff}$ and the depopulation of Zr~I levels is less intense when using $n$ than that when using $n_{eff}$. Increasing $\sigma_{ph}(n)$ by a factor of 10 causes the abundance determined from Zr~I lines to increase by 0.01~dex. A great effect (-0.09~dex) is obtained when $\sigma_{ph}$ is reduced by a factor of 10, but it leads to a reduction in the departures from LTE for Zr~I and, hence, to poorer agreement between Zr~I and Zr~II. When using $n_{eff}$, increasing the photoionization cross sections $\sigma_{ph}(n_{eff})\times$10 does not lead to a noticeable change in the departures from LTE, while the abundance determined from Zr~I lines with $\sigma_{ph}(n_{eff})$/10 decreases by 0.01~dex.

Varying the photoionization cross sections affects very weakly the Zr~II level populations, because this is the dominant ionization stage. The change in the abundance determined from Zr~II lines is everywhere smaller than 0.01~dex.
We also calculated the change in abundance when varying the rate of collisions with electrons: $C_{ij}(e)\times$10 for $k_H$ = 0.0 and $k_H$ = 0.1+ (Table~\ref{delta-nltecorr}). The influence of collisions with electrons on the statistical equilibrium of zirconium must be strongest in the absence of hydrogen collisions. With $k_H$ = 0.0, the difference log$\varepsilon_{Zr}(C_{ij}(e)\times10 - C_{ij}(e))$ ranges between $-0.01$ and $-0.05$~dex for different Zr~I and Zr~II lines, except the Zr~II 5112~$\ang$ line for which this difference is $-0.08$~dex. However, when the hydrogen collisions are taken into account, the influence of an uncertainty in $C_{ij}(e)$ on the departure from LTE decreases, because $C_{ij}(H)$ is an order of magnitude larger than $C_{ij}(e)$ for the transitions between levels with a small energy difference at which the collisional processes are particularly important. For example, for the standard case of $k_H$ = 0.1+, the difference log$\varepsilon_{Zr}(C_{ij}(e)\times10 - C_{ij}(e))$ is from 0 to -0.03~dex. The sign of the effect is the same for the lines of both ionization stages and, hence, the difference between the mean abundances changes by no more than 0.01~dex.

We see that the errors in the zirconium abundance due to the uncertainty in the photoionization cross sections and in the rate of collisions with electrons are smaller than the error in the mean abundance attributable to the spread in values obtained from different lines ($\sigma$ = 0.07~dex).

Thus, allowance for the departures from LTE makes it possible to reconcile, within the error limits, the zirconium abundances determined from the lines of the two ionization stages and we obtain the mean zirconium abundance log$\varepsilon_{Zr}$ = 2.63$\pm$0.07 (for $k_H$ = 0.1+).

\subsection{Comparison with other studies}

The LTE zirconium abundance on the Sun has been determined previously by several authors. One of the first studies is the work by Biemont and Grevesse \cite{biemont}, where the authors used 34 Zr~I lines and 24 Zr~II lines and the method of equivalent widths. The mean abundances derived from the lines of each ionization stage agree well between themselves and with the meteorite abundance, but with large errors: log$\varepsilon_{Zr}$(Zr~I) = 2.57 $\pm$0.21 and log$\varepsilon_{Zr}$(Zr~II) = 2.56 $\pm$0.14. In our view, the large errors are attributable to the difficulty of allowance for the blending of most of the Zr~I and Zr~II lines used.

We pay particular attention to the study by Ljung et al \cite{ljung}. They determined the zirconium abundance from seven Zr~II lines (Table~\ref{zr2lines}) by the synthetic spectrum method using the solar spectrum in intensities and a three-dimensional model solar atmosphere. Subsequently, the equivalent widths of Zr~II lines were calculated from the theoretical spectrum and the zirconium abundances were also determined using the MARCS \cite{gustaf} and Holweger– Muller \cite{hol-mu} models from equivalent widths.

For comparison, we made additional calculations with the Holweger-Muller model and the solar spectrum in intensities from the disk center from the atlas by Brault and Testerman \cite{brault}. From the private communication by Asplund \cite{asplund-rep}, we know that Ljung et~al. \cite{ljung} used the microturbulent velocity V$_{mic}$ = 1.0\,\kms. Therefore, we also took V$_{mic}$ = 1.0 km s$^{-1}$ for our calculations. The results are presented in Table~\ref{our-lju}. The equivalent widths of Zr~II lines that we measured in the intensity spectrum turned out to agree with those from Ljung et~al.\cite{ljung} within 1 m$\ang$, except for the Zr~II 4442.992 and 4050.320~$\ang$ lines for which the difference between our values of W$\lambda$ and those determined by Ljung et~al.\cite{ljung} is 1.5 and 2.9 m$\ang$, respectively. In this case, our abundances are systematically higher than those obtained by Ljung et~al. \cite{ljung} by a value from 0.07 to 0.11~dex for all of the lines, except Zr~II 4050.320~$\ang$ for which the difference of our values and those obtained by Ljung et~al. \cite{ljung} is $\Delta\varepsilon_{Zr} = -0.04$~dex. The mean abundance from our determinations, log$\varepsilon_{Zr}$ = 2.68 $\pm$ 0.06, is higher by 0.05~dex that the value obtained by Ljung et~al.\cite{ljung}, log$\varepsilon_{Zr}$ = 2.63$\pm$0.02.

\begin{table}
\caption{Equivalent width and abundance comparisons between this study and Ljung et~al. \cite{ljung} for the selected lines of Zr~II. Everywhere, the calculations were performed with the Holweger \& Muller \cite{hol-mu} solar model atmosphere.}\label{our-lju}
\bigskip
\begin{tabular}{ccccc}
\hline
$\lambda^L$, $\ang$ $\rule{0pt}{15pt}$& ~~W$_{\lambda}^V$, m$\ang$~~& ~~W$_{\lambda}^{L}$, m$\ang$ ~~& ~~log$\varepsilon_{Zr}^V$~~ & ~~log$\varepsilon_{Zr}^{L}$~~ \\
\hline
\hline
4050.320                             &  19.1    &  22.0   & 2.59     & 2.63     \\
4208.980                             &  42.5    &  42.6   & 2.66     & 2.59     \\
4258.041                             &  23.8    &  23.4   & 2.69     & 2.63     \\
4442.992                             &  21.6    &  20.4   & 2.71     & 2.62     \\
4496.962                             &  31.7    &  31.5   & 2.70     & 2.65     \\
5112.270                             &  8.6     &  7.8    & 2.76     & 2.65     \\
\hline
\multicolumn{5}{l}{\small $^L$ - data from \cite{ljung}$\rule{0pt}{15pt}$}      \\
\multicolumn{5}{l}{\small $^V$ - this work}                          \\
\end{tabular}
\end{table}

The difference between our values of W$\lambda$ and those of Ljung et~al.\cite{ljung} can, in part, be explained by the subtleties of line profile fitting. For example, according to the private communication by Asplund \cite{asplund-rep}, they corrected their equivalent width of the Zr~II 5112~$\ang$ line and obtained a new value of W$\lambda$(5112) = 8.5 m$\ang$, which agrees with our determination. However, the zirconium abundance determined from this line changed by only 0.01~dex: the value from Ljung et~al. \cite{ljung} is log$\varepsilon_{Zr}$ = 2.64, while our value obtained from this line is log$\varepsilon_{Zr}$ = 2.76. Thus, the difference in line equivalent widths cannot explain the difference in derived abundances.

The cause of the discrepancies in zirconium abundance between our determinations and those of Ljung et~al. \cite{ljung} remains unclear. It may be related to differences in the techniques of calculations.

\section{Non-LTE effects depending on stellar parameters}\label{NLTE_effects}

Here, we investigate the influence of non-LTE effects on Zr~I and Zr~II lines for a small grid of model atmospheres with effective temperatures T$_{eff}$ = 5500 and 6000 K, surface gravities logg = 2.0 and 4.0, and metallicities [M/H] = $-3$, $-2$, $-1$, 0. Everywhere, microturbulence velocity is V$_{mic}$ = 1.0\,\kms. In these calculations, we took into account the collisions with neutral hydrogen atoms in both permitted and forbidden transitions with $k_H$ = 0.1. The results are presented in Table~\ref{nlte-cor} as differences between the non-LTE and LTE abundances: $\Delta_{\rm NLTE}$ = log$\varepsilon_{\rm NLTE}$ - log$\varepsilon_{\rm LTE}$. Hereafter, $\Delta_{\rm NLTE}$ is referred to as non-LTE correction. The non-LTE corrections are positive in all cases for both Zr~I and Zr~II lines.

\begin{table}
\caption{The NLTE abundance corrections for the Zr~I and Zr~II lines depending on stellar parameters.}
\label{nlte-cor}
\bigskip
\begin{tabular}{l|cc|ccccccccc}
\hline
\raisebox{-7pt}[0pt][0pt]{T$_{eff}$/logg/[M/H]}& \multicolumn{2}{c|}{Zr~I}& \multicolumn{9}{c}{Zr~II} \\
   & 4241 & 4687 & 3479.3 & 3505 & 3551 & 4050 & 4208 & 4258 & 4442 & 4496 & 5112 \\
\hline
\hline
~~5500/2.0/0         & 0.32 & 0.31 & 0.01 & 0.00 & 0.03 & 0.03 & 0.01 & 0.01 & 0.01 & 0.01 & 0.06  \\
~~5500/2.0/-1        &      &      & 0.16 & 0.08 & 0.07 & 0.14 & 0.11 & 0.07 & 0.10 & 0.10 & 0.18  \\
~~5500/2.0/-2        &      &      & 0.34 & 0.19 & 0.15 & 0.16 & 0.18 & 0.10 & 0.20 & 0.12 &       \\
~~5500/2.0/-3        &      &      & 0.22 & 0.19 & 0.19 & 0.18 & 0.19 &      &      &      &       \\
~~5500/4.0/0         & 0.31 & 0.33 & 0.06 & 0.02 & 0.03 & 0.00 & 0.05 & 0.02 & 0.04 & 0.03 & 0.06  \\
~~5500/4.0/-1        &      & 0.49 & 0.14 & 0.05 & 0.04 & 0.06 & 0.08 & 0.03 & 0.09 & 0.04 & 0.16  \\
~~5500/4.0/-2        &      &      & 0.12 & 0.07 & 0.07 & 0.10 & 0.12 & 0.07 &      & 0.08 &       \\
~~5500/4.0/-3        &      &      & 0.11 & 0.10 & 0.10 &      &      &      &      &      &       \\
~~6000/2.0/0         & 0.31 & 0.30 & 0.01 & 0.01 & 0.05 & 0.06 & 0.03 & 0.02 & 0.02 & 0.03 & 0.08  \\
~~6000/2.0/-1        &      &      & 0.21 & 0.13 & 0.07 & 0.13 & 0.13 & 0.07 & 0.09 & 0.10 & 0.13  \\
~~6000/2.0/-2        &      &      & 0.29 & 0.14 & 0.14 & 0.11 & 0.14 & 0.07 & 0.11 & 0.08 & 0.14  \\
~~6000/2.0/-3        &      &      & 0.20 & 0.15 & 0.15 &      &      &      &      &      &       \\
~~6000/4.0/0         & 0.31 & 0.32 & 0.08 & 0.03 & 0.03 & 0.05 & 0.05 & 0.03 & 0.04 & 0.04 & 0.08  \\
~~6000/4.0/-1        &      &      & 0.17 & 0.07 & 0.06 & 0.08 & 0.10 & 0.05 & 0.12 & 0.06 &       \\
~~6000/4.0/-2        &      &      & 0.13 & 0.10 & 0.10 &      & 0.12 &      &      &      &       \\
~~6000/4.0/-3        &      &      & 0.12 &      & 0.12 &      &      &      &      &      &       \\
\hline
\end{tabular}
\end{table}

\begin{figure}
\includegraphics{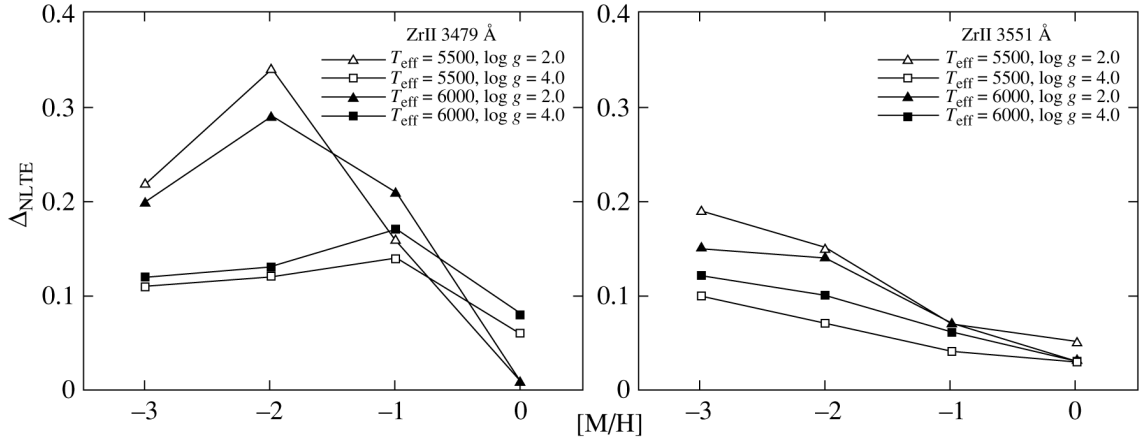}
\caption{The NLTE abundance corrections for Zr~II 3479 and 3551~$\ang$ depending on stellar parameters.}
\label{nlte-dif}
\end{figure}

The non-LTE effects are strong for the Zr~I lines in the solar metallicity models, with $\Delta_{\rm NLTE}$ reaching to 0.33~dex for T$_{eff}$ = 5500~K and logg = 4.0. We do not give data for the lower metallicity models because of very small equivalent widths of the Zr~I lines ($W_\lambda <$ 1~m$\ang$). An exception is Zr~I 4687~$\ang$ in the model 5500/4.0/$-1$.

%Since are very weak, they can be observed mainly only at [M/H] = 0 and  at a . We consider a line to be unmeasurable if its We conclude that should be taken into account for Zr~I lines, because they are great: 

%Since the Zr~II lines are stronger, the dependence of non-LTE effects on atmospheric parameters can be traced by them. For clarity, 
Figure~\ref{nlte-dif} shows the non-LTE corrections for the Zr~II 3479 and 3551~$\ang$ lines depending on metallicity, effective temperature, and surface gravity. These lines can be detected in low-metallicity stars, down to [M/H] = $-3$. As can be seen from Fig.~\ref{nlte-dif} and Table~\ref{nlte-cor}, the abundance corrections are small for Zr~II lines at solar metallicity: $\Delta_{\rm NLTE} <$ 0.08~dex. However, it should be remembered that disregarding the departures from LTE introduces a systematic error that underestimates the derived abundance. The departures from LTE for the Zr~II 3551~$\ang$ line increase with decreasing metallicity. For Zr~II 3479~$\ang$ in the models with logg = 2.0, the non-LTE corrections increase from 0.01~dex at [M/H] = 0 up to $\sim$0.34~dex at [M/H] = $-2$. This is explained by an increase in the ultraviolet flux, on the one hand, and by a decrease in the number density of free electrons, on the other hand, and, as a result, by an increase in the radiative rate and a decrease in the collisional rate. As the metallicity decreases further, the non-LTE correction drops to 0.2~dex at [M/H] = $-3$ due to shifting the line formation depth to deeper atmospheric layers where collisions are efficient to establish thermodynamic equilibrium. 
%an increase in the depth of line formation in the atmosphere where the density is higher than that in the higher layers. 
A similar behavior is also observed for the Zr~II 3479~$\ang$ line in 
the models with T$_{eff}$ = 5500~K and logg = 4.0 and T$_{eff}$ = 6000 K and logg = 4.0. The non-LTE correction increases from 0.06~dex and 0.08~dex at [M/H] = 0 to 0.14~dex and 0.17~dex at [M/H] = $-1$ and then decreases to 0.11~dex and 0.12~dex at [M/H] = $-3$. The departures from LTE for Zr~II lines are weakly sensitive to changes in effective temperature, which can also be seen from Fig.~\ref{nlte-dif}.

\section{Conclusions}\label{conclusions}

We constructed a model zirconium atom that includes 63 levels of Zr~I , 247 levels of Zr~II, and the ground Zr~III state. We showed that the Zr~I levels are underpopulated relative to their LTE populations due to ultraviolet overionization, while the excited Zr~II levels are overpopulated because of radiative pumping from the ground state and lower excited levels. Non-LTE leads to weakening both Zr~I and Zr~II lines relative to their LTE strengths and positive non-LTE abundance corrections.

%Our calculations for a grid of model atmospheres showed that 
The departures from LTE are strong for the Zr~I lines in the solar metallicity models, with $\Delta_{\rm NLTE}$ at the level of 0.3~dex. For most Zr~II lines
in the models with logg = 4.0, 
%at  did not exceed the errors in the zirconium abundance, 
$\Delta_{\rm NLTE} \leq$ 0.12~dex. An exception is Zr~II 3479 and 5112~$\ang$, with non-LTE corrections of 0.17 and 0.16~dex, respectively. The departures from LTE for Zr~II are predicted to be stronger in giant stars (logg = 2). For example, $\Delta_{\rm NLTE}$(Zr~II 3479~$\ang$) = 0.34~dex in the model with T$_{eff}$ = 5500~K, logg = 2.0, and [M/H] = $-2$. 
All the stars studied by Mashonkina et~al. \cite{yzrce} at the LTE assumption are dwarfs or subgiants, with logg $\ge 3.12$. Taking the non-LTE effects for Zr~II into account would lead to an up to 0.2~dex higher Zr overabundance relative to Ba than that found in \cite{yzrce}. Thus, we support our earlier conclusion about the growth of the Zr/Ba ratio towards lower barium abundance. 

%They should be taken into account when the Zr~II lines are analyzed. 
%The dependence of non-LTE effects on atmospheric parameters cannot be traced for Zr~I lines, because they are very weak and become unmeasurable even at [M/H] = $-1$.

From our analysis of Zr~I and Zr~II lines in the solar spectrum, we found that the non-LTE approach led to better agreement between the abundances derived from the lines of the two ionization stages than the LTE one. We could not constrain the scaling factor $k_H$ to the classical Drawin formalism due to the large spread in abundances derived from different lines. Nevertheless, we dwelt on the non-LTE case with allowance made for the hydrogen collisions in both permitted and forbidden transitions with $k_H$ = 0.1. The absolute zirconium abundance in the solar atmosphere (averaged over Zr~I and Zr~II lines) is log$\varepsilon_{Zr}$ = 2.63$\pm$0.07.

{\it Acknowledgments.}
This work was supported by the Russian Foundation for Basic Research (project 08-02-92203-GFEN-a), the Russian Federal Agency for Science and Innovations (project 02.740.11.0247), and the Swiss National Science Foundation (SCOPES project IZ73Z0-128180/1).

\newpage

{\bf Erratum: Formation of Zr I and II lines under non-LTE conditions of stellar atmospheres [Astron. Letters, 36, 664 (2010)]}

In the published paper, an outdated value of $E_{ion}$(Zr~I) = 6.84~eV was employed for an ionization energy of the neutral zirconium, instead of $E_{ion}$(Zr~I) = 6.63~eV measured by Hackett et~al. (Hackett, P. A., Humphries M. R., Mitchell S. A., \& Rayner D. M., 1986, J. Chem. Phys. 85, 3194)
 and recommended by the NIST atomic data base (http://physics.nist.gov/PhysRefData). 
The calculations with $E_{ion}$(Zr~I) = 6.63~eV lead to lower number density of Zr~I and, therefore, higher element abundance derived from the Zr~I lines. For the Sun, such an abundance correction amounts to 0.18~dex independent of the line-formation assumptions used. The change in $E_{ion}$(Zr~I) only weakly affects the departure coefficients of the important Zr~I levels in the solar atmosphere, i.e., by no more than 1.5/3.2\,\%\ inside $\log\tau_{5000} = -1/-2$. This is because the ratio $J_\nu/B_\nu(T)$ changes only a little when a wavelength of the ultraviolet ionizing photons increases by approximately 60\,\AA. The change in $E_{ion}$(Zr~I) does not affect the results for Zr~II.  

The corrections to the published paper are as follows.

1. Table~1, Table~3, and Fig.~7: the Zr~I based abundances have to be increased by 0.18~dex.

2. Abstract, Sect.~3.2, last paragraph, and Sect.~5, last paragraph. The revised abundance difference ($\eps{}$(Zr~I) -- $\eps{}$(Zr~II)) is $-0.10$~dex at the LTE assumption and $+0.10$ up to $+0.15$~dex in the non-LTE calculations depending on the treatment of hydrogenic collisions. The calculations with $E_{ion}$(Zr~I) = 6.63~eV support our earlier conclusion that the Zr~I/Zr~II ionization equilibrium in the solar atmosphere cannot be established under the LTE conditions.
The magnitude of the departures from LTE for Zr~I can be overestimated 
due to missing high-excitation levels of Zr~I in the model atom. 

The authors thank Nicolas Grevesse for picking our mistake up.


\begin{thebibliography}{99}
\bibitem{anders}
E. Anders \& N. Grevesse,  Geoch. Cosmochim Acta, \textbf{53}, 197 (1989) 
\bibitem{aoki05}
W. Aoki, S. Honda, T.C. Beers, et al., 2005, ApJ, \textbf{632}, 611 
\bibitem{arlandini}
C. Arlandini, F. K$\rm\ddot{a}$ppeler, K, Wisshak et al., Astrophys. J., \textbf{525}, 886-990 (1999) 
\bibitem{asplund}
M. Asplund, N. Grevesse, \& A. J. Sauval, ASP Conf. Ser., \textbf{336}, 25 (2005) 
\bibitem{asplund-rep}
M. Asplund, private communication (2009) 
\bibitem{brault}
J. Brault \& L. Testerman, Preliminary Kitt Peak Photoelectric Atlas, Nat. Sol. Obs, Tucson (1972) 
\bibitem{belyaev}
A. K. Belyaev, J. Grosser, J. Hahne, \& T. Menzel, Phys. Rev. A, \textbf{60}, 2151 (1999) 
\bibitem{bel-bark}
A. K. Belyaev \& P. Barklem, Phys. Rev. A, \textbf{68}, 062703 (2003) 
\bibitem{biemont}
E. Bi\'{e}mont \& N. Grevesse, Astrophys. J., \textbf{248}, 867 (1981)
\bibitem{detail}
K. Butler \& J. Giddings, Newsletter on the analysis of astronomical spectra No. 9, University of London (1985) 
\bibitem{drav}
H. W. Drawin, Z. Phys., \textbf{164}, 513 (1961) 
\bibitem{francois2007}
P. Fran\c{c}ois, E. Depagne, V. Hill, et al. 2007, Astron. Astrophys, \textbf{476}, 935 
\bibitem{mafag}
K. Fuhrmann, M. Pfeiffer, C. Frank, et al., Astron. Astrophys., \textbf{323}, 909 (1997) 
\bibitem{gustaf}
B. Gustafsson, R. A. Bell, K. Eriksson, \& A. Nordlund, Astron. Astrophys, \textbf{42}, 407 (1975) 
%\bibitem{HHMR86}
%Hackett, P. A., Humphries M. R., Mitchell S. A., \& Rayner D. M., 1986, J. Chem. Phys. 85, 3194
\bibitem{hol-mu}
H. Holweger \& E. A. M$\rm\ddot{u}$ller, Sol. Phys., \textbf{39}, 19 (1974) 
\bibitem{kappeler}
F. K${\rm\ddot{a}}$ppeler, H. Beer and K. Wisshak, Rep. Prog. Phys. \textbf{52}, 954 (1989) 
\bibitem{kupka1999}
F. Kupka, N. Piskunov, T. A. Ryabchikova, H. C. Stempels, \& W. W. Weiss,  Astron. Astrophys, \textbf{138}, 119 (1999) 
\bibitem{sol-at}
R. L. Kurucz, I. Furenlid, J. Brault, and L. Testerman, Solar Flux Atlas from 296 to 1300 nm, Nat. Solar Obs., Sunspot, New Mexico (1984) 
\bibitem{kur}
R.L. Kurucz, CD-Roms No. 18, 19 (1994) 
\bibitem{lodders}
K. Lodders, Astrophys.J, \textbf{591}, 1220 - 1247 (2003) 
\bibitem{Lod-zr-ba}
K. Lodders, H. Palme, \& H.-P. Gail, Landolt-B$\rm\ddot{o}$rnstein, New Ser., Astron. Astrophys., Springer Verlang, Berlin (2009) 
\bibitem{ljung}
G. Ljung, H. Nilsson, M. Asplund, \& S. Johansson,  Astron. Astrophys, \textbf{456}, 1181 (2006) 
\bibitem{malch}
G. Malcheva, K. Blagoev, R. Mayo, et al., MNRAS, \textbf{367}, 754 (2006) 
\bibitem{yzrce}
L. I. Mashonkina, A. B. Vinogradova, D. A. Ptitsyn, et al., Astron. Rep., \textbf {51}, 11 (2007)
\bibitem{reetz}
J. K. Reetz, Diploma Thesis, Universit$\rm\ddot{a}$t M$\rm\ddot{u}$nchen (1991) 
\bibitem{vanreg}
H. van Regemorter, Astrophys. J., \textbf{136}, 906 (1962) 
\bibitem{sh}
W. Steenbock \& H. Holweger,  Astron. Astrophys, \textbf{130}, 319  (1984) 
\bibitem{takeda}
Y. Takeda, PASJ, \textbf{46}, 53 (1994)

\end{thebibliography}
\end{document}